\documentclass[twocolumn, nobibnotes, aps, prl, superscriptaddress]{revtex4-1}

\usepackage{graphicx}
\usepackage{float}
\usepackage{braket}
\usepackage{amsmath,amssymb}

\usepackage[colorlinks=true, allcolors=blue]{hyperref}

\newcommand{\IAB}{I_\mathrm{AB}}
\newcommand{\SAB}{S_\mathrm{AB}}
\newcommand{\SA}{S_\mathrm{A}}
\newcommand{\SB}{S_\mathrm{B}}
\newcommand{\A}{\mathrm{A}}
\newcommand{\B}{\mathrm{B}}
\def\beq{\begin{equation}}
\def\eeq{\end{equation}}

\global\long\def\ket#1{\left| #1\right\rangle }%
\global\long\def\bra#1{\left\langle #1 \right|}%
\global\long\def\kket#1{\ket{\ket{#1}}}%
\global\long\def\bbra#1{\bra{\bra{#1}}}%
\global\long\def\bbrakket#1#2{\left\langle \left\langle #1 \middle\|  #2 \right\rangle \right\rangle }%
\global\long\def\abs#1{\left|#1\right|}%

\DeclareMathOperator{\tr}{tr}

\DeclareMathOperator{\sign}{sign}

\begin{document}

\title{Quantum Noise as a Symmetry-Breaking Field}%

\author{Beatriz~C.~Dias}
\affiliation{Max Planck Institute for the Physics of Complex Systems, N\"{o}thnitzer Str. 38, 01187 Dresden, Germany}
\affiliation{CeFEMA, Instituto Superior T\'{e}cnico, Universidade de Lisboa, Av. Rovisco Pais, 1049-001 Lisboa, Portugal}

\author{Domagoj Perkovi\'{c}}
\affiliation{Max Planck Institute for the Physics of Complex Systems, N\"{o}thnitzer Str. 38, 01187 Dresden, Germany}
\affiliation{Department of Physics, Cavendish Laboratory, JJ Thomson Avenue, Cambridge,
CB3 0HE, UK.}

\author{Masudul~Haque}
\affiliation{Institut f\"ur Theoretische Physik, Technische Universit\"at
Dresden, 01062 Dresden, Germany}
\affiliation{Department of Theoretical Physics, Maynooth University, Co.\ Kildare, Ireland}
\affiliation{Max Planck Institute for the Physics of Complex Systems, N\"{o}thnitzer Str. 38, 01187 Dresden, Germany}
\author{Pedro~Ribeiro}
\affiliation{CeFEMA, Instituto Superior T\'{e}cnico, Universidade de Lisboa, Av. Rovisco Pais, 1049-001 Lisboa, Portugal}
\affiliation{Beijing Computational Science Research Center, Beijing 100084, China}
\author{Paul~A.~McClarty}
\affiliation{Max Planck Institute for the Physics of Complex Systems, N\"{o}thnitzer Str. 38, 01187 Dresden, Germany}

\begin{abstract}
We investigate the effect of quantum noise on the measurement-induced quantum phase transition in monitored random quantum circuits. Using the efficient simulability of random Clifford circuits, we find that the transition is broadened into a crossover and that the phase diagram as a function of projective measurements and noise exhibits several distinct regimes. We show that a mapping to a classical statistical mechanics problem accounts for the main features of the random circuit phase diagram. The bulk noise maps to an explicit permutation symmetry breaking coupling; this symmetry is spontaneously broken when the noise is switched off. These results have implications for the realization of entanglement transitions in noisy quantum circuits.
\end{abstract}

\maketitle
%\tableofcontents

From the point of view of entanglement entropy,
ground states of many-body quantum systems with short-range interactions mostly fall into two classes:
those with area law entanglement, where entanglement entropy scales
as $L^{D-1}$, with $L$ the system's linear extent and $D$ the dimension,
or critical phases with entanglement entropy scaling as $L^{D-1}\ln L$.
% Typical continuous quantum phase transitions between two phases are area to area law transitions with an intervening critical point as in the transverse field Ising model, or from an area law to a critical phase as in metal-insulator transitions. 
%
Out of equilibrium, it also becomes possible to generate volume law entanglement and to induce continuous area to volume law transitions. Examples have been found in disordered
spin chains between thermalizing and localized phases \cite{RevModPhys.91.021001,nandkishore2015many,Parameswaran_2018,alet2018} and
in systems subject to measurements. One example
of a measurement-induced entanglement transition is a spin chain under
local random unitary gates and projective measurements \cite{PhysRevB.98.205136,PhysRevB.99.224307, VasseurPotterLudwig_PRB2019_HolographicTensorNetworks, PhysRevX.9.031009,Li2019b,PhysRevLett.125.030505,PhysRevX.10.041020, BaoChoiAltman_PRB2020,  PhysRevResearch.2.013022, VasseurLudwig_PRB2020_classicalSMmapping, ZabaloGullansGopalakrishnanHusePixley_PRB2020,  SierantTurkeshi_PRL2022_multifractality, ZabaloGullansVasseurLudwigGopalakrishnanHusePixley_PRL2022}.
In the absence of measurements, repeated application of gates to an
initial product of local pure states leads to a linear build-up of
the entanglement entropy \cite{NahumRuhmanVijayHaah_PRX2017, KeyserlingkRakovszkyPollmannSondhi_PRX2018, NahumVijayHaah_PRX2018} saturating at the Page value \cite{Page1993}.
The effect of carrying out local projective measurements with rate
$p$, in the limit where $p$ is large, is to suppress the spread
of quantum information completely so that the entanglement obeys an
area law. In between there is a continuous quantum phase transition
at which the volume law coefficient vanishes \cite{PhysRevB.98.205136,PhysRevB.99.224307,PhysRevX.9.031009}.
%
% This phenomenon has been investigated in local random Clifford and Haar distributed unitary circuits in the steady state and the dynamics \cite{Li2019b,PhysRevLett.125.030505,PhysRevX.10.041020,  PhysRevResearch.2.013022, VasseurLudwig_PRB2020_classicalSMmapping}.
%
% Similar physics has been explored in fully connected random circuits and trees \cite{PRXQuantum.2.010352}.
%
Measurement-induced entanglement transitions of different sorts have by 
now been observed in free fermion systems \cite{Alberton_Buchhold_Diehl_PRL2021, Buchhold_Altland_Diehl_PRX2021, BaoChoiAltman_AnnPhys2021, Ladewig_Buchhold_Diehl_PRR2022},
in fully connected random circuits \cite{PRXQuantum.2.010352}, in systems with weak measurements \cite{PhysRevB.100.064204}, continuous monitoring \cite{Alberton_Buchhold_Diehl_PRL2021, PhysRevLett.125.210602,PhysRevB.102.054302, TurkeshiDalmonteFazioSchiro_PRB2022_nonhermquasiparticles,  MinoguchiRablBuchhold_SciPost2022}, and long-range interactions \cite{MinatoSugimotoKuwaharaSaito_PRL2022_longrange, Turkeshi_Fazio_Dalmonte_SciPost2022_longrange, HashizumeBentsenDaley_PRR2022, BlockBaoChoiAltmanYao_PRL2022_longrange, MuellerDiehlBuchhold_PRL2022,  SierantTurkeshiDalmonteFazio_Quantum2022_DissipativeFloquet}.
The weight of evidence is that these continuous phase transitions are
much like their equilibrium counterparts with emergent conformal invariance
and associated critical exponents, though a complete field theory
description is lacking.

\begin{figure}[t!]
\centering
\includegraphics[width=\columnwidth]{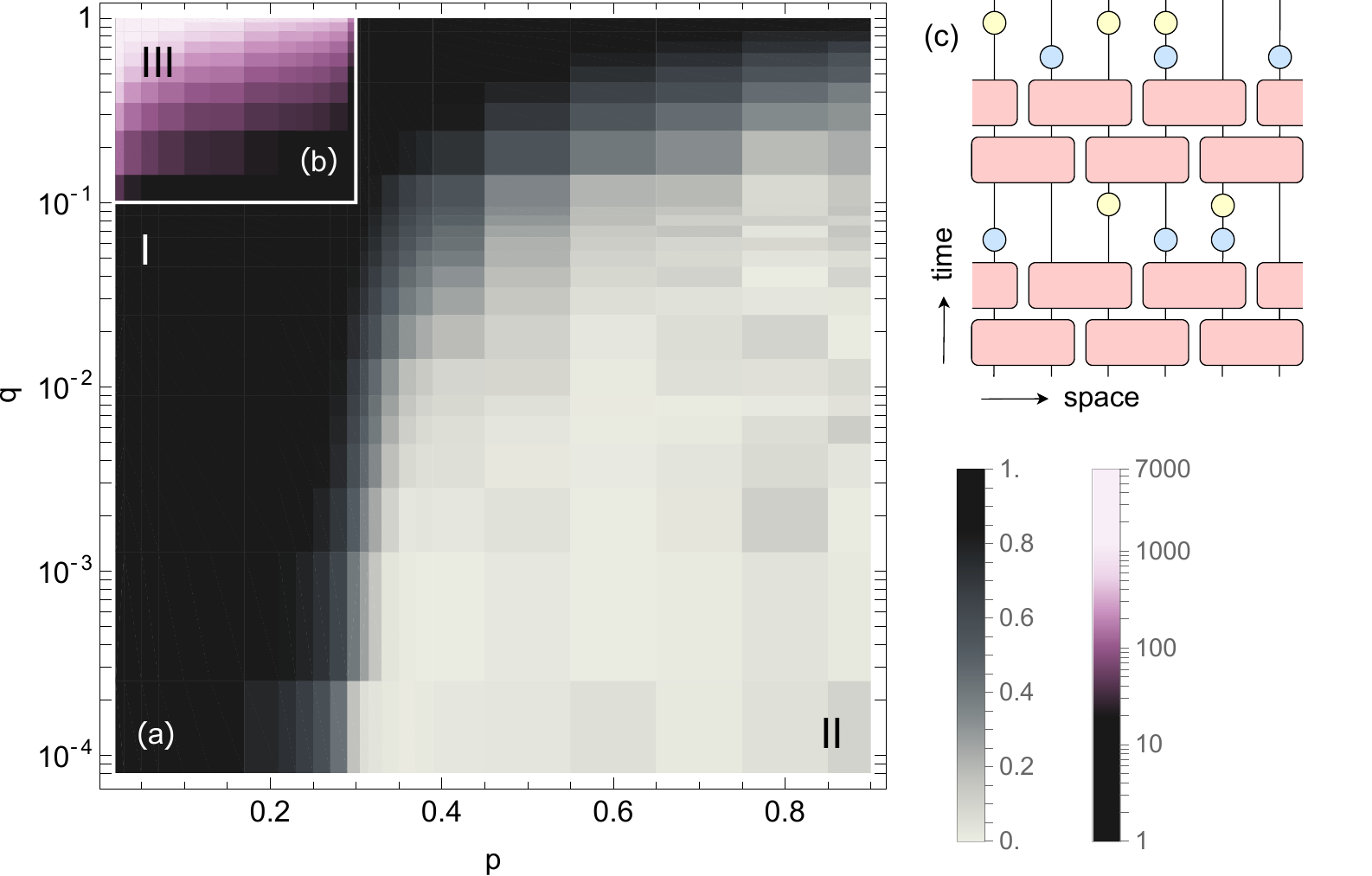}
\caption{\label{fig:1} 
(a) $|\IAB(L, p, q) - \IAB(L, p, q = 0)| /\IAB(L, p, q = 0)$ and (b) $|\IAB(L, p, q) - f(L, p, q)|/\IAB(L, p, q)$ for a system with $L = 500$. For each $p$, $f(L, p, q)=b(L,p) \log(1/q) + c(L,p)$ is fitted to $\IAB(L, p, q)$ with $q$ in region I, where $\IAB \sim \log(q)$. The three marked regions correspond to: $\IAB \sim \log(1/q)$ (I), $\IAB \sim q^0$ (II) and $\IAB \sim \exp(-q)$ (III).
(c) A 1D qubit chain is evolved by applying random uniformly distributed Clifford gates (red blocks) to neighboring pairs of sites in a brickwall pattern. After each Clifford bilayer, an un monitored measurement is performed on each site with probability $q$ (blue circles), and then a monitored measurement is performed on each site with probability $p$ (yellow circles).}
\end{figure}

Since any physical system is always coupled to an environment, it is natural to ask what becomes of these measurement-induced entanglement
transitions in the generic case where the system is noisy.  Refs.~\cite{BaoChoiAltman_AnnPhys2021, WeinsteinBaoAltman_PRL2022_open} made the case that quantum noise breaks down a symmetry present in the noise-free case and Ref.~\cite{WeinsteinBaoAltman_PRL2022_open}  studied a model with noise at the boundary.  
%
% with a phase characterized by an $L^{1/3}$ logarithmic negativity. 
%
In this
paper, we address the effect of noise in the bulk
with projective measurements applied at rate $p$ and with some additional
noise rate $q$. We find that the known transition
at $p=p_{c}$, $q=0$ is broadened and the finite $p,q$ phase diagram
exhibits three quantitatively distinct regimes separated by crossovers,  Fig.~\ref{fig:1}(a,b). At small $q$ this is reminiscent
of a phase diagram where $q=0$ displays spontaneous symmetry breaking
and where $q$ plays the role of an explicit symmetry breaking field. 
We show, through a replica construction, that this picture can be
made precise: the quantum noise appears as a new effective coupling in an effective classical model that explicitly breaks a replica permutation
symmetry. Using the classical picture, we can account for the main features of the random circuit phase diagram. In the following,
we introduce the protocol, discuss our numerical findings and motivate
a classical model that allows for their qualitative understanding.

\begin{figure}[b]
\centering
\includegraphics[width=\linewidth]{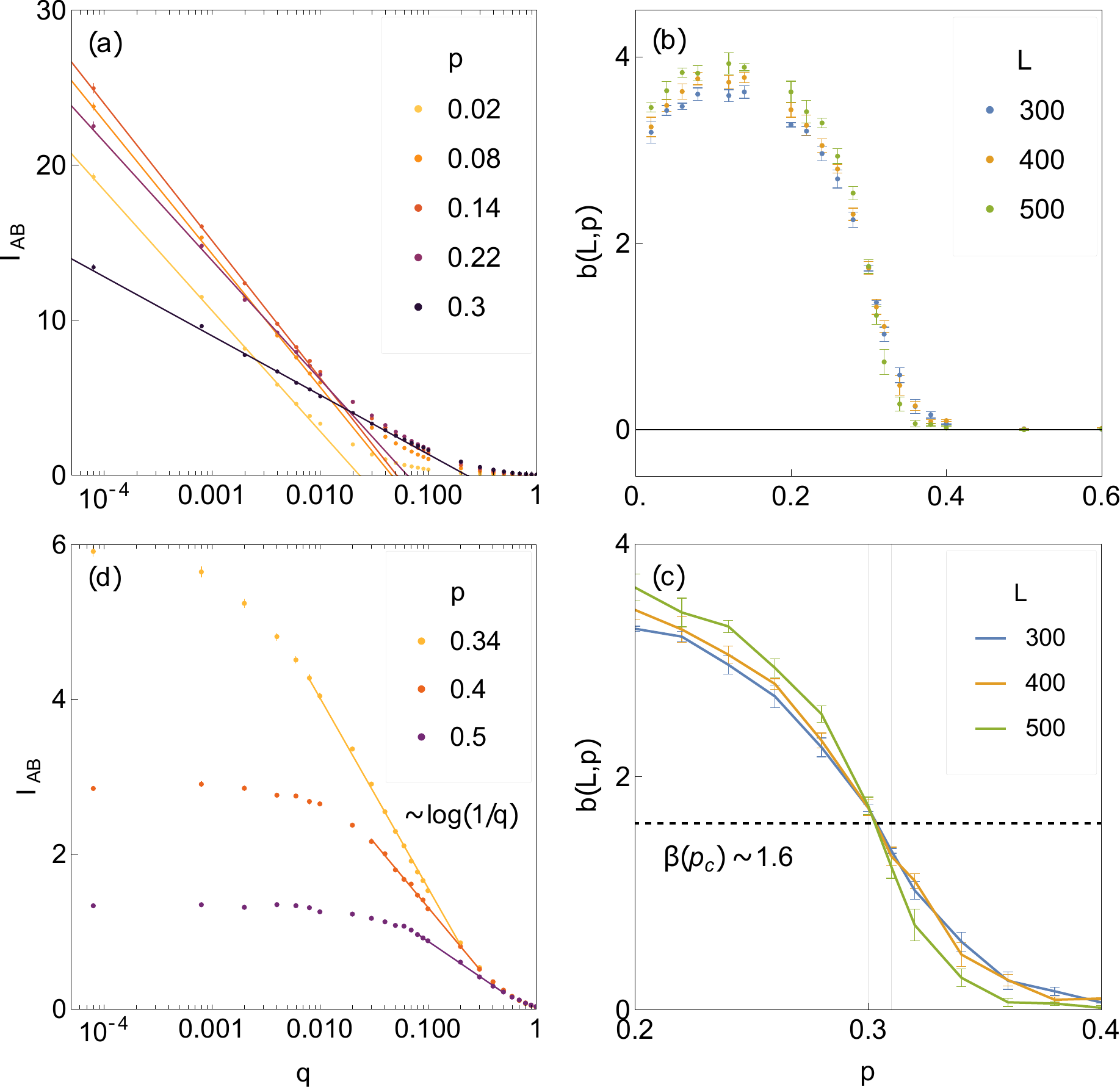}
\caption{\label{fig:2}
(a) Mutual information, $\IAB(q)$, for a system with $L = 500$ and for  $p\in\{0.02, 0.08, 0.14,0.22,0.3\}$. Solid lines are $f(L, p) = b(L,p) \log(1/q) + c(L,p)$ fits to the points with $q=0.00008, 0.0008, 0.002, 0.004$. 
(b) Parameter $b(L,p)$ of the fit in (a). 
(c) Zoom in of (b): the critical point of the transition at $q=0$ is $p_c \in ]0.30,0.31[$. The horizontal dashed line is $\beta(p_c) \sim 1.6$ \cite{Li2019b} in $\SA(L,p,q=0)=\alpha(p)L+\beta(p)\log L+\gamma(p)$. 
(d) Mutual information, $\IAB(q)$, for a system with $L = 500$ and for $p\in\{0.34, 0.4, 0.5\}$. Solid lines are $b(L,p) \log(1/q) + c(L,p)$ fits to the points in region I of Fig.~\ref{fig:1}. One can see that $\IAB(q) \sim q^0$ for small $q$.
}
\end{figure}

%%%%%%%%%%%%%%%%%%%%%%%%%%%%%%%%%%%%%%%%%%%%%%%%%%%%%%%%%%%%%%%%%%%
%%%%%%%% CIRCUIT
%%%%%%%%%%%%%%%%%%%%%%%%%%%%%%%%%%%%%%%%%%%%%%%%%%%%%%%%%%%%%%%%%%%

\textit{Circuit and Observables --} We consider a circuit model
consisting of a chain of $L$ qudits with local Hilbert space dimension
$d$. Each single discrete time step of the circuit  is composed of four layers (Fig.~\ref{fig:1}(c)). The first two
layers consist of local random unitary gates $\prod_{r}u_{2r,2r+1}$
followed by $\prod_{r}u_{2r-1,2r}$ in a brick wall pattern. After
the random unitary bilayer, unmonitored measurements are performed
on each site, $r$, with probability $q$. These take the state $\rho$
to the mixed state $\sum_{a}M_{r}^{a\dagger}\rho M_{r}^{a}$, where
$M_{r}^{a}=\vert r,a\rangle\langle r,a\vert$ for outcomes $a=1,\ldots,d$.
The final layer performs monitored (projective) measurements with probability $p$,
taking the mixed state $\rho$ to $M_{r}^{a\dagger}\rho M_{r}^{a}/||M_{r}^{a\dagger}\rho M_{r}^{a}||$
with probability $p_{a}=||M_{r}^{a\dagger}\rho M_{r}^{a}||$.

We characterize the system using the entropy of subsystem A, $\SA=-{\rm Tr}_{\A}(\rho_{\A}\log\rho_{\A})$,
with $\rho_{\A}={\rm Tr}_{\mathrm{\text{B}}}\rho$ the reduced density
matrix of subsystem A, obtained by tracing out its complement, $\text{B}$.
Since the overall state of $\A\cup\B$ is no longer pure, we also
consider the mutual information, $\IAB=S_{\mathrm{A}}+S_{\mathrm{B}}-\SAB$,
with $\SAB$ the entanglement entropy of the subsystem $\A\cup\B$.
We note that $\IAB=2S_{\mathrm{A}}$ when $q=0$ as $\SAB$ vanishes
in this case.

%%%%%%%%%%%%%%%%%%%%%%%%%%%%%%%%%%%%%%%%%%%%%%%%%%%%%%%%%%%%%%%%%%%
%%%%%%%% OBSERVABLES
%%%%%%%%%%%%%%%%%%%%%%%%%%%%%%%%%%%%%%%%%%%%%%%%%%%%%%%%%%%%%%%%%%%

\textit{Numerical Phase Diagram}
$-$ In order to achieve system sizes up to a few hundred sites, we
set $d=2$ and make use of the efficient simulability of Clifford
circuits with both above-mentioned single-site measurements, i.e.
of stabilizer circuits \cite{PhysRevA.70.052328}. As uniformly sampled
gates within the stabilizer group constitute a $t$-design of order
$t=3$ \cite{10.5555/3179439.3179447,PhysRevA.96.062336}, the second Renyi entropy of local random local
Clifford gates with monitored and unmonitored measurements reproduce
the same result as their local Haar-distributed counterpart \cite{Berg2020,Li2019b,PhysRevX.10.041020}.
For convenience we take an entanglement cut dividing the chain exactly
in two. 

We now analyze the numerical results of Fig.~\ref{fig:1}(a,b),
starting from the previously known $p=0$ line. For $p=q=0$, the
circuit rapidly loses memory of the initial state and the entanglement
entropy grows linearly saturating at a value that is proportional
to $L$. For $q=0$, $p=1$, there is a projective measurement at
each site for every time step so the state is a product state with
short range entanglement. There is a transition in the entanglement
at $0.30 < p_{c} < 0.31$, $q=0$ \footnote{The reported value in Ref.~\cite{PhysRevB.98.205136} is approximately
half of this value i.e. $p_{c}=0.16$ because their circuit has one
measurement layer for each (even or odd) unitary layer.}.  The entanglement entropy obeys an area law for $p>p_{c}$ and varies
as $\SA(L,p_c,q=0)=\beta(p_c) \log L$ at criticality. Empirically, for
$p<p_{c}$, there is a volume law with a logarithmic
correction, $\SA(L,p,q=0)=\alpha(p)L+\beta(p)\log L+\gamma(p)$, with $\lim_{p \rightarrow p_c^-} \beta(p) \neq 0$
and $\lim_{p \rightarrow p_c^+} \beta(p) = 0$, and $\alpha(p)=(p_{c}-p)^{\nu}\Theta(p_{c}-p)$
with $\nu\approx1.3$ \cite{Li2019b}.

Now we turn to the finite $q$ behavior. 
%
% as determined from the simulated Clifford circuit. 
%
For $p=0$, $q\neq 0$, the state asymptotes to the infinite temperature mixed state essentially because the unmonitored measurements proliferate branches of measurement outcomes in the density matrix leading to rapid decoherence. As the gates are uniformly distributed, the circuit does not distinguish between measurement outcomes so the diagonal entries in the measurement basis become equal. 

For finite $p$ and $q$, the steady state density matrix is nontrivial as there is competition between branch proliferation generated by $q$ and branch reduction driven by $p$. We find three distinct finite $p$, $q$ regimes as indicated in the phase diagram of Fig.~\ref{fig:1}.   The $q=0$ transition is broadened into a crossover at finite $q$. In addition, the mutual information obeys an area law for all finite $q$. For $p>p_c$ and small $q$, the mutual information $\IAB$ is $q$ independent and equal to the area law entanglement $\SAB$ for $q=0$. In the region I of Fig.~\ref{fig:1} (spanning $p<p_c$) and small $q$ all the way to $p>p_c$ and large $q$ there is a logarithmic regime $\IAB(L,p,q) = b(L,p) \ln (1/q)+c(L,p)$. This scaling is shown in Fig.~\ref{fig:2}(a) for $p<p_c$ and above the saturation regime for $p>p_c$ in panel (d). At the transition point itself, the finite size cuts off the logarithmic scaling at $b(L,p_c) \ln L$. We find that the coefficients $b(L,p_c)$ and $\beta(p_c)$ are compatible, as shown in Fig.~\ref{fig:2}(c). This has the important implication that the underlying correlation length grows as a power law $\xi\sim 1/q^\nu$ with exponent $\nu=1$. For $p<p_c$, the logarithmic divergence is cut off by the volume law entanglement implying logarithmic growth of the correlation length. A third scaling regime is observed for small $p$ and $q$ close to $1$ where $\IAB$ falls off exponentially in $q$ \cite{SM}.   

%%%%%%%%%%%%%%%%%%%%%%%%%%%%%%%%%%%%%%%%%%%%%%%%%%%%%%%%%%%%%%%%%%%
%%%%%%%% CLASSICAL MODEL
%%%%%%%%%%%%%%%%%%%%%%%%%%%%%%%%%%%%%%%%%%%%%%%%%%%%%%%%%%%%%%%%%%%

{\it Mapping to a Classical Statistical Mechanics Model} $-$  The smearing of the phase transition into a crossover at finite $q$ is reminiscent of a classical equilibrium statistical mechanics model where $q$ explicitly breaks the symmetry that is spontaneously broken by tuning $p$. We now make this connection more concrete by mapping the $p,q$ model to a classical statistical mechanics problem building on work from Ref.~\cite{ZhouNahum_PRB2019_statmechmodel} for random unitary circuits and Ref.~\cite{VasseurLudwig_PRB2020_classicalSMmapping} for the $q=0$ model. It will turn out that this mapping accounts for many other features of the $p,q$ phase diagram. 

For this section, we generalize the qubit ($2$-states per site) model considered above to a model of qudits ($d$-states per site). We consider the $n$th R\'{e}nyi entropy $ S_{A}^{\left(n\right)}  =1/(1-n)\ln\left[{\rm Tr}_{A}\left(\rho_{A}^{n}\right)\right]$, 
averaged over Haar random unitaries $\left\{ U\right\}$ within each layer of the circuit and measurement outcomes, ${\bf m}$, taken to include both monitored and un-monitored measurements
\beq
\bar{S}_{A}^{\left(n\right)}  =\int dU\sum_{{\bf m}}\ \frac{1}{1-n}\ln\left[\frac{{\rm Tr}_{A}\left(\rho_{A,{\bf m}}^{n}\right)}{{\rm Tr}\left(\rho_{{\bf m}}\right)^{n}}\right]{\rm Tr}\left(\rho_{{\bf m}}\right). 
\eeq
We introduce the swap operator
%\begin{align}
$
\Sigma_{A}^{\left(n\right)}  =\sum_{\left[i\right]}\ket{i_{g^n_{\text{cyc}}\left(1\right)}^{A}i_{1}^{B}, \ldots, i_{g^n_{\text{cyc}}\left(n\right)}^{A}i_{n}^{B}}\bra{i_{1}^{A}i_{1}^{B}, \ldots, i_{n}^{A}i_{n}^{B}},
$
%\end{align}
which  performw a cyclic permutation of $n$ replicas on  $A$ while leaving B sites invariant. We also use the identity $\ln x=\lim_{k\to0}\frac{x^{k}-1}{k}$ to rewrite the logarithms, introducing an additional replica index $k$, giving
\begin{align}
\bar{S}_{A}^{\left(n,k\right)} &  =\frac{1}{k(1-n)}{\rm Tr}\left\{ \left[ \Sigma_{n,A}^{\otimes k}\otimes1-1^{\otimes\left(nk+1\right)}\right] \right. \nonumber \\ & \left. \times \left[\int dU\sum_{m}\rho_{m}^{\otimes\left(nk+1\right)}\right]\right\}. 
\label{eq:SAnk}
\end{align}
In this form one may integrate over the random unitaries and the measurement outcomes.  Details are spelled out in \cite{SM}.  Here we  report the outcome of the averaging. 

\begin{figure}[t!]
\centering
\includegraphics[width=\columnwidth] {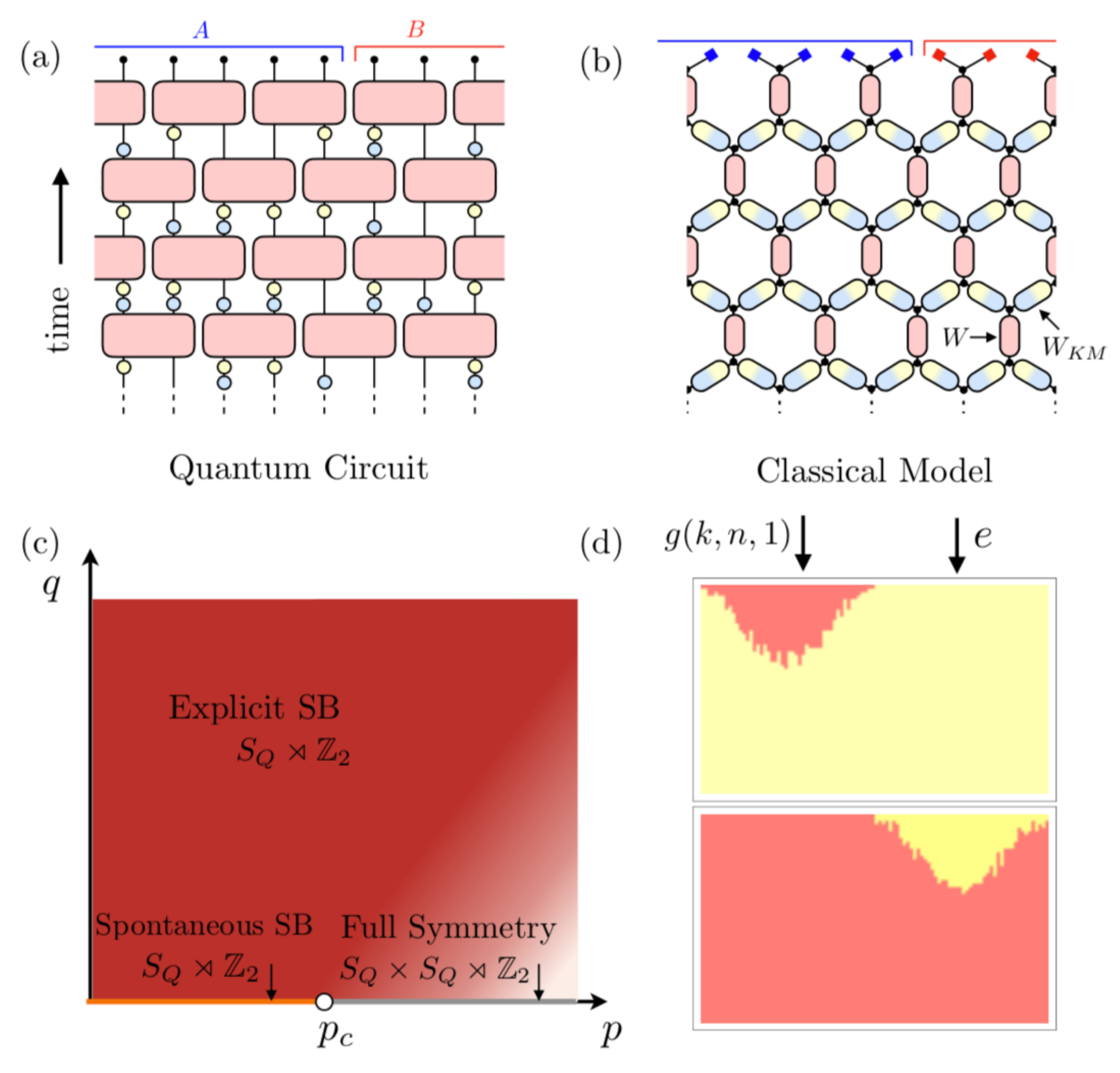}
\caption{\label{fig:3} 
Top: mapping from the quantum random circuit (a) with both projective and un-monitored measurements to a classical model with permutation degrees of freedom living on the sites of an anisotropic honeycomb lattice (b). The vertical bonds carry weights coming from the random unitaries and the half-shaded bonds carry weights depending on $p$ and $q$. Panel (c) is the schematic phase diagram arising from the classical model. Panel (d) shows two domain wall configurations with pinning on the upper boundary to $g(k,n,1)$ on the left and $e$ on the right. In the spontaneously broken phase both configurations can arise with equal weight whereas in the explicitly broken phase the state is biased towards the upper plot with $e$ across the system.
}
\end{figure}

It turns out that $\bar{S}_{A}^{\left(n,k\right)}$ is related to the free energy cost of a domain wall in a classical statistical mechanics model \cite{ZhouNahum_PRB2019_statmechmodel}. This classical lattice model has degrees of freedom $g_i$ on each site, which are elements of the permutation group $\mathcal{S}_Q$ of $Q\equiv kn+1$ objects. The partition sum is:
\begin{align}
\mathcal{Z}_{A} & = \sum_{\left\{ g_i \in S_Q\right\} } \left[ \prod_{\langle i,j \rangle \in \mathcal{E}_{\rm V}} W(g_i g_j^{-1})  \prod_{\langle i,j \rangle \in \mathcal{E}_{\rm ZZ}} W_{KM}(g_i, g_j) \right. \nonumber \\ \times & \left. \prod_{i\in \partial_0} d^{\abs{g_i}} \prod_{i\in \partial_T \cap B } d^{\abs{g_i}} \prod_{i\in \partial_T \cap A } d^{\abs{g(n,k,1)^{-1} g_i}} \right].
\label{eq:ZA}
\end{align}
%\beq
%\mathcal{Z}_{\emptyset} = \sum_{g_i \in \mathcal{S}_Q} \prod_{\langle i,j \rangle \in \mathcal{E}_{\rm ZZ}} W_{KM}(g_i,g_j) \prod_{\langle i,j \rangle \in \mathcal{E}_{\rm V}} W(g^{-1}_i g_j). 
%\label{eq:Z0}
%\eeq
The anisotropic couplings are between nearest neighbors on a honeycomb lattice (Fig.~\ref{fig:3}(b)). With $i,j$ as neighboring honeycomb sites and $\mathcal{E}_{\rm V}$ denoting the vertical bonds %(see Fig.~\ref{fig:honeycomb}) 
and $\mathcal{E}_{\rm ZZ}$ the zigzag bonds, the $W(g^{-1}_ig_j)$ act on the $\mathcal{E}_{\rm V}$ bonds. These are Weingarten functions originating from the average over unitaries. On the $\mathcal{E}_{\rm ZZ}$ bonds, the weights for $q=0$ are $
\left. W_{KM}(g,g')\right\vert_{q=0} = (1-p)^Q d^{\abs{ g^{-1}g'}} + p^Q d^Q$. Here $\vert g \vert$  means the number of cycles in permutation $g$. The second line in Eq.~\ref{eq:ZA} consists of the boundary conditions that bias permutation degrees of freedom towards the trivial permutation on the early and late time boundaries except along $A$ where they are fixed to permutation $g(n,k,1)$ \cite{SM}. Crucially, boundaries aside, the model has a global $S_Q \times S_Q \rtimes \mathbb{Z}_2$ symmetry. From Eq.~\ref{eq:ZA} and weights one sees that, for large $p$, the local degrees of freedom fluctuate wildly while, for small $p$, permutations on neighboring bonds tend to lock together. In fact, there is a critical $p$, analogous to temperature, below which there is long-range order that spontaneously breaks the symmetry down to $S_Q \rtimes \mathbb{Z}_2$. 

One may now understand features of the two phases from the classical model as the averaged replicated R\'{e}nyi entropy is
$\bar{S}_{A}^{\left(n\right)} = \frac{n}{1-n} \frac{\mathcal{Z}_{A}-\mathcal{Z}_{\emptyset}}{Q-1}$
where $\mathcal{Z}_{A}$ is defined in Eq.~\ref{eq:ZA} and $\mathcal{Z}_{\emptyset}$ has homogeneous boundary conditions corresponding to $A\rightarrow \emptyset$. In fact, the averaged R\'{e}nyi entropy is the free energy cost of a domain wall with fixed permutations (Fig~\ref{fig:3}(d) unbiased between the two configurations shown). Evidently, this costs $O(L)$ in the ordered phase and $O(L^0)$ in the paramagnetic phase corresponding to the volume and area law phases respectively. 

One may extend this calculation to include the un-monitored measurements. The weights $W_{KM}$  then become 
%\begin{widetext}
\begin{align}
& W_{KM}(g_i,g_j)=(1-p)^Q \left[ (1-q)^{Q}d^{\abs{gg'^{-1}}} \right. \nonumber \\ &  \left. +\sum_{l=1}^{Q}q^{l}(1-q)^{Q-l}\sum_{\{r_{1},\ldots,r_{l}\}\in B_{l,Q}}d^{|gg'^{-1}h_{r_{1}}^{t_{1}}\ldots h_{r_{l}}^{t_{l}}|} \right] + p^Q d^Q
\label{eq:Wpq}
\end{align}
%\end{widetext}
where $B_{l,Q}$ is the set of subsets of ${1,2,\ldots,Q}$ with $l$ elements and where $h_{r}=(g^{-1}(r)\ r)$ is a two-cycle that permutes replica index $r$ and $g^{-1}(r)$ and
$   t_{k}= 1/2+{\rm sign}(\vert gg'^{-1}\vert-\vert gg'^{-1}h_{r_k}\vert )/2, 
$ 
is either 0 or 1. Despite the notational complexity, the $h_r$ insertions have a relatively simple effect. That is to say if, in the permutation
$gg'^{-1}$, the elements $g^{-1}(r)$ and $r$ belong to different
cycles, then $gg'^{-1}h_{r_{k}}^{t_{k}}$ joins the cycles to which
$g^{-1}(k)$ and $k$ belong and $\abs{gg'^{-1}}-|gg'^{-1}h_{r_{k}}|=1$.
If the elements $g^{-1}(k)$ and $k$ already belong to the same cycle, then $\abs{gg'^{-1}}-|gg'^{-1}h_{r_{k}}|=-1$ and $t_k=0$, such that $gg'^{-1}h_{r_{k}}^{t_{k}}=gg'^{-1}$. Overall Eq.~\ref{eq:Wpq} reduces to the $q=0$ weight quoted above. 

{\it Comparison of classical picture and quantum model} $-$ We can now interpret the random circuit results from the point of view of the classical model. The primary effect of introducing quantum noise is an explicit breaking of the symmetry of the model down to $S_Q \rtimes \mathbb{Z}_2$ to which, we recall, it is spontaneously broken when $q=0$. This can be seen more transparently by inspecting the large $d$, small $q$ limit. In this limit, the $q=0$ model reduces to a $Q!$ state Potts model. Expanding Eq.~\ref{eq:Wpq} linearly in $q$ and collecting the largest powers of $d$ we obtain \cite{SM}
\beq
W_{KM}(g_i, g_j) = d^Q \Big( (1-p)^Q \big[ (1-q) + q|g_j|_1 \big] \delta_{g_i,g_j} + p^Q  \Big)
\label{eq:Wpqexpand}
\eeq
where $|g|_1$ denotes the number of 1-cycles in $g$. Thus the leading effect of $q$ is to pick out the permutation with the largest number of cycles -- the trivial permutation from all $Q!$ states equivalent to applying directly a symmetry-breaking field that biases the system towards the $e$ permutation with an extensive contribution to the free energy 
(Fig.~\ref{fig:3}(d), now biased towards the upper configuration). 

Thus, for finite $p$, quantum noise drives the system to the infinite temperature fixed point of the classical model.  This accounts for the most prominent feature of the phase diagram: the smearing of the transition into a crossover for $q\neq 0$. For $p=0$, the system instead flows to the infinite-field fixed point corresponding to the maximally mixed state.

Turning to the observables, we note that $S_{AB}$ vanishes for $q=0$ but exhibits a volume law for $q\neq 0$. From the point of view of the classical model, when $q=0$, $S_{AB}$ is the free energy difference of a system with boundary degrees of freedom pinned to a nontrivial permutation across the entire system and a system with a boundary pinned to the trivial permutation (Fig.~\ref{fig:3}(d)). But, for $q=0$, the free energy is indifferent to the precise pinning field so the free energy difference must vanish. This ceases to be the case when $q\neq 0$ but $q$ distinguishes the trivial permutation leading to an $O(L)$ free energy difference as observed in the quantum model. A similar argument allows us to understand why $I_{AB}$ has an area law for $q\neq 0$: the domain wall cost is a constant $\alpha$ times $L$ in $S_{AB}$ which is cancelled off by the two $\alpha L/2$ contributions in $S_A$ and $S_B$. 

Finally the classical model provides some insight into the $\log q$ dependence of the mutual information. In the vicinity of the $q=1$ line, one may carry out the analogue of a high temperature expansion with Boltzmann weights $\exp(c \log q)$. Since $q$ is close to $1$, the appropriate expansion parameter is $\log q$. Then for $p< p_c$, the only further scale that enters the problem is $L$ but until this cutoff is reached there is nothing to interfere with the $\log q$ growth from the high temperature expansion. 

{\it Discussion and Conclusion} $-$ We have introduced and studied a random circuit model subjected to both projective and un-monitored measurements.  We have shown that,  at finite $q$,  the transition known for projective-only measurements ($q=0$) is washed out and the phase diagram is reminiscent of that of a finite temperature ferromagnet with $q$ playing the role of an applied field. We have made this connection precise through a mapping to a classical statistical mechanics model with finite replica index $Q$ (the quantum model corresponding to replica index $Q\rightarrow 1$). In the classical model, $q$ explicitly breaks down the global $S_Q \times S_Q \rtimes \mathbb{Z}_2$ permutation symmetry of the model to $S_Q \rtimes \mathbb{Z}_2$. Thus, quantum noise (originating, e.g., from coupling to a bath) is a relevant operator driving the system to the infinite temperature fixed point. We have seen that various features of the quantum model follow from the classical statistical mechanics analogue. It is amusing to note that the passage from pure to mixed states by switching on $q$ is reflected in the classical model as a field that affects the response of the system but not the state space. Indeed, in the classical model the difference between the pure and mixed states is only apparent through boundary conditions. 

% What is the broader significance of this result? 
%
All quantum systems are subject to quantum noise coming from the environment. Our result suggests that, in a realistic setting, the entanglement transition driven by monitored measurements will be smeared out by the environment.  In addition, our work shows that nonequilibrium mixed state dynamics in Hilbert space in the presence of a quantum bath can be viewed as breaking a delicate symmetry that is only present for pure states.  It would be of considerable interest to devise a model with an entanglement transition in the presence of bulk noise \cite{PhysRevLett.125.030505}. Whether this is possible or not has a bearing on the experimental simulability of these entanglement transitions as well as implications for error corrected quantum computers \cite{PhysRevA.62.062311}. 

\begin{acknowledgments}
BD and PR acknowledge MPI PKS where part of this work was carried out.
\end{acknowledgments}

\bibliography{paper_measurements}

\newpage % \. \newpage

%%%%%%%%%%%%%%%%%%%%%%%%%%%%%%%%%%%%%%%%%%%%%%%%%%%%%%%
% Counter resetting for supplementaries
\setcounter{page}{1} \renewcommand{\thepage}{S\arabic{page}}

\setcounter{figure}{0}   \renewcommand{\thefigure}{S\arabic{figure}}

\setcounter{equation}{0} \renewcommand{\theequation}{S.\arabic{equation}}

\setcounter{table}{0} \renewcommand{\thetable}{S.\arabic{table}}

\setcounter{section}{0} \renewcommand{\thesection}{S.\Roman{section}}

\renewcommand{\thesubsection}{S.\Roman{section}.\Alph{subsection}}

\makeatletter
\renewcommand*{\p@subsection}{}
\makeatother

\renewcommand{\thesubsubsection}{S.\Roman{section}.\Alph{subsection}-\arabic{subsubsection}}

\makeatletter
\renewcommand*{\p@subsubsection}{}  % referring to subsubsections
\makeatother

%%%%%%%%%%%%%%%%%%%%%%%%%%%%%%%%%%%%%%%%%%%%%%%%%%%%%%%

\newpage \;  \newpage

\begin{widetext}

\begin{center}
\underline{\large {\em Supplementary Material:} Quantum Noise as a Symmetry-Breaking Field}
\end{center}

\section{Exponential regime (region III)}

In the main text we reported that the small $p$, large $q$ region (region III) displays an exponential decrease of $I_{AB}$ with $q$. In Figure \ref{fig:expfit} we present numerical data displaying this behavior.

\begin{figure}[H]
\centering
\includegraphics[width=0.5\columnwidth]{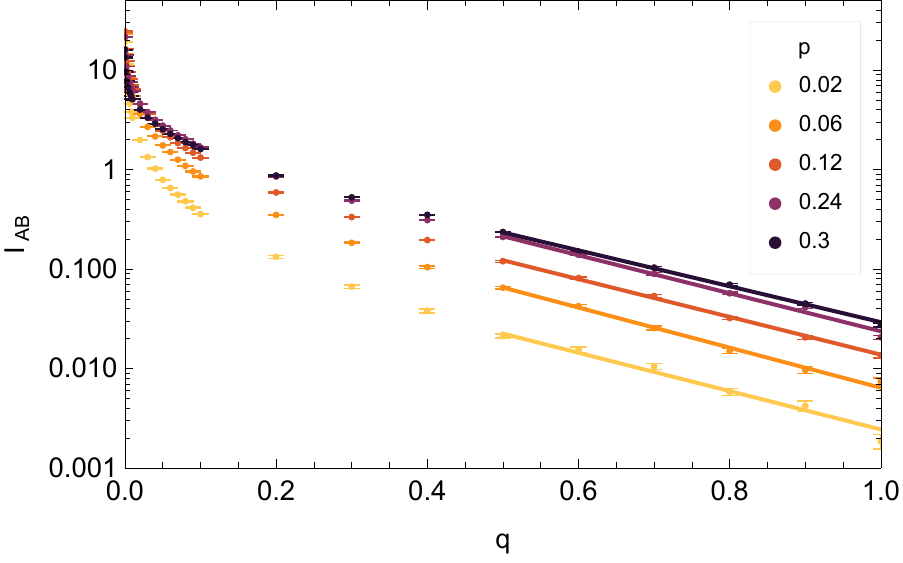}
\caption{\label{fig:expfit} Mutual information, $\IAB(q)$, for a system with $L = 500$ and for $p\in\{0.02, 0.06, 0.16,0.22,0.3\}$. Solid lines are $d(L,p) \exp( - e(L,p) q )$ fits to points with large $q$ ($q=0.5,0.6,0.7,0.8,0.9,1.0$) belonging to the region III of Fig.~1 in the main text, where $\IAB(q) \sim \exp(-q)$.}
\end{figure}

\section{Classical Mapping for the Mixed State Circuit}

As discussed in the main text, Ref.~\cite{VasseurLudwig_PRB2020_classicalSMmapping} (building on work from Ref.~\cite{ZhouNahum_PRB2019_statmechmodel}) presents a mapping of the suitably replicated quantum random circuit model with projective measurements to a classical lattice model at finite temperature. Within the classical statistical mechanics model the volume to area entanglement transition is an order-disorder transition involving a broken permutation symmetry in the volume law phase. The entanglement scaling follows from the free energy cost of a domain wall. 

Here we review this work and generalize it to the case with both projective measurements and un-monitored measurements. 

\subsection{Replicas}

We wish to compute the n\textsuperscript{th} Rényi entropy 
\begin{align}
S_{A}^{\left(n\right)} & =\frac{1}{1-n}\ln\left[\tr_{A}\left(\rho_{A}^{n}\right)\right]
\end{align}
for the circuit model with Haar-random unitaries acting on $d$ dimensional local degrees of freedom (qudits), projective measurements, $M$, and un-monitored measurements, $K$. This is to be done by averaging over circuit disorder and measurement outcomes as follows:
\begin{align}
S_{A}^{\left(n\right)} & =\int dU\sum_{m}\ \frac{1}{1-n}\ln\left[\frac{\tr_{A}\left(\rho_{A,m}^{n}\right)}{\tr\left(\rho_{m}\right)^{n}}\right]\tr\left(\rho_{m}\right) \nonumber\\
 & =\int dU\sum_{m}\ \frac{1}{1-n}\left\{ \ln\left[\tr_{A}\left(\rho_{A,m}^{n}\right)\right]-\ln\left[\tr\left(\rho_{m}\right)^{n}\right]\right\} \tr\left(\rho_{m}\right).
\end{align}
Here the density matrix for some choice of measurements (indicated in the subscript) is
\begin{align}
\rho_{m} & =\sum_{\mu}K_{\mu_{T}}M_{m_{T}}U_{T}...K_{\mu_{1}}M_{m_{1}}U_{1}\rho_{0}U_{1}^{\dagger}M_{m_{1}}^{\dagger}K_{\mu_{1}}^{\dagger}\ldots U_{T}^{\dagger}M_{m_{T}}^{\dagger}K_{\mu_{T}}^{\dagger},
\end{align}
where $T$ is the circuit depth, $L$ the circuit length and 
\begin{align}
M_{0} & =\sqrt{1-p},\\
M_{i=1,\ldots,d} & =p\ket i\bra i,\\
K_{0} & =\sqrt{1-q},\\
K_{i=1,\ldots,d} & =q\ket i\bra i.
\end{align}
The integral is performed with the standard Haar measure. 
Using $\ln x=\lim_{k\to0}\frac{x^{k}-1}{k}$, we get a replica index $k$ 
\begin{align}
\\
S_{A}^{\left(n\right)} & =\lim_{k\to0}S_{A}^{\left(n,k\right)}
= \lim_{k\to0}\frac{1}{1-n}\frac{1}{k}\int dU\sum_{m}\left\{ \left[\tr_{A}\left(\rho_{A,m}^{n}\right)\right]^{k}-\left[\tr\left(\rho_{m}\right)\right]^{nk}\right\} \tr\left(\rho_{m}\right).
\end{align}
Defining the swap operator 
\begin{align}
\Sigma_{A}^{\left(n\right)} & =\sum_{\left[i\right]}\ket{i_{g^n_{\text{cyc}}\left(1\right)}^{A}i_{1}^{B}, \ldots, i_{g^n_{\text{cyc}}\left(n\right)}^{A}i_{n}^{B}}\bra{i_{1}^{A}i_{1}^{B}, \ldots, i_{n}^{A}i_{n}^{B}},
\end{align}
where $g^n_{\text{cyc}}$ is a cyclic permutation of $n$ replicas, and $i_{r}^{A(B)}$
runs over a complete basis of replica $r$ of subsystem $A$($B$),
we can write 
\begin{align}
S_{A}^{\left(n,k\right)} & =\frac{1}{1-n}\frac{1}{k}\tr\left\{ \left[\Sigma_{A}^{\left(n\right)}{}^{\otimes k}\otimes1-1^{\otimes\left(nk+1\right)}\right]\left[\int dU\sum_{m}\rho_{m}^{\otimes\left(nk+1\right)}\right]\right\}. 
\end{align}
Or, in vectorial notation and defining $Q \equiv n k + 1$,
\begin{align}
S_{A}^{\left(n,k\right)} & =\frac{1}{1-n}\frac{1}{k}\left[\bbra{\Sigma_{A}^{\left(n,k\right)}}-\bbra 1\right]\kket{\bar{\rho}_{Q}},
\label{eq:SAnk}
\end{align}
with
\begin{align}
\kket{\bar{\rho}_{Q}} &= \int dU\sum_{m}\kket{\rho_{m}}^{\otimes Q},\\
\kket{\Sigma_{A}^{\left(n\right)}} &=\left(\sum_{\left[i\right]}\ket{i_{g^n_{\text{cyc}}\left(1\right)}^{A}i_{1}^{B},\ldots,i_{g^n_{\text{cyc}}\left(n\right)}^{A}i_{n}^{B}}\otimes\ket{i_{1}^{A}i_{1}^{B},\ldots,i_{n}^{A}i_{n}^{B}}\right)^{\otimes k}\otimes\left(\sum_{\left[i_{Q}\right]}\ket{i_{Q}^{A}i_{Q}^{B}}\otimes\ket{i_{Q}^{A}i_{Q}^{B}}\right),
\label{eq:SgA}\\
\kket{1_{Q}} &= \left(\sum_{i}\ket{i^{A}i^{B}}\otimes\ket{i^{A}i^{B}}\right)^{\otimes Q}.
\end{align}

We use Eq.~\ref{eq:SAnk} to introduce a classical partition function as indicated in the main text. In particular, we define 
\begin{align}
\bar{S}_{A}^{\left(n,k\right)} = \frac{n}{1-n} \frac{\mathcal{Z}_{A}-\mathcal{Z}_{\emptyset}}{Q-1}
\end{align}
where 
\begin{align}
\mathcal{Z}_{A} = \bbrakket{ \Sigma_{A}^{\left(n\right)} }{ \bar{\rho}_{Q}}
\label{eq:Z_A}
\end{align}
and 
\begin{align}
\mathcal{Z}_{\emptyset} = \bbrakket{ 1 }{ \bar{\rho}_{Q}}.
\label{eq:Z_emptyset}
\end{align}

\subsection{Averaging Unitaries}

The averaged replicated density matrix is given by 
\begin{align}
\kket{\bar{\rho}_{Q}}= & \int dU\sum_{m}\kket{\rho_{m}}^{\otimes Q},
\end{align}
where 
\begin{align}
\kket{\rho_{m}} & =\left(\sum_{\mu_{T}}K_{\mu_{T}}\otimes K_{\mu_{T}}^{*}\right)\left(M_{m_{T}}\otimes M_{m_{T}}^{*}\right)\left(U_{T}\otimes U_{T}^{*}\right)...\left(\sum_{\mu_{1}}K_{\mu_{1}}\otimes K_{\mu_{1}}^{*}\right)\left(M_{m_{1}}\otimes M_{m_{1}}^{*}\right)\left(U_{1}\otimes U_{1}^{*}\right)\kket{\rho_{0}}.
\end{align}
Let us analyze the operators averaged over replica space: 
\begin{align}
\mathcal{U} & =\int dU\left(U\otimes U^{*}\right)^{\otimes Q},\\
\mathcal{M} & =\sum_{m}\left(M_{m}\otimes M_{m}^{*}\right)^{\otimes Q},\\
\mathcal{K} & =\left(\sum_{\mu}K_{\mu}\otimes K_{\mu}^{*}\right)^{\otimes Q}.
\end{align}
For $U$ acting on two qubits we have 
\begin{align}
\mathcal{U}_{12} & =\int dU\left(U_{12}\otimes U_{12}^{*}\right)^{\otimes Q}  =\sum_{g_{1}g_{2} \in S_Q}W\left(g_{1}^{-1}g_{2}\right)\kket{g_{1}}\kket{g_{1}}\bbra{g_{2}}\bbra{g_{2}}
\label{eq:Weingarten}
\end{align}
where $g_1$ and $g_2$ are permutations of $Q$ elements, and $\kket g$ is a state on the replica-copy
space of one of the qubits, given by 
\begin{align}
\kket g & =\sum_{\left[i\right]}\kket{i_{g\left(1\right)},\ldots,i_{g\left(Q\right)};i_{1},\ldots,i_{Q}}.
\label{eq:g}
\end{align}
A schematic representation of this relation is given in Fig.\ref{fig:Weingarten} The coefficient $W(g)$ is a Weingarten function on permutation $g$. In short, the average over unitaries leads to the emergence of classical permutation variables - each pairwise unitary operator gives two permutation degrees of freedom. Thus, on the lattice, the brick wall pattern of unitaries maps to a lattice model with permutation degrees of freedom on two sites for each unitary. 

\begin{figure}[H]
\centering
\includegraphics[width=0.75\columnwidth]%{2022-08-01_fig1_W.pdf}
{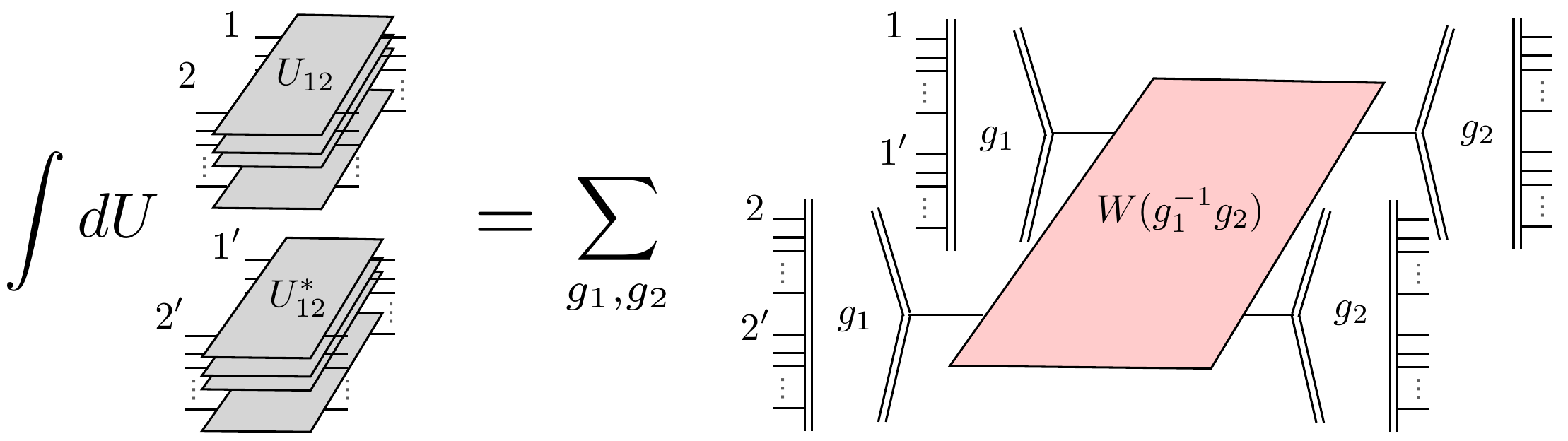}
\caption{\label{fig:Weingarten} Sketch of Eq.~\ref{eq:Weingarten}. The integration over the Haar distributed unities, $U_{1,2}$, acting individual on $Q$ replicas of qudits 1 and 2 and their complex conjugate $U_{1,2}^*$, acting on copy-space on $Q$ sets of qudits 1' and 2', is replaced by sums over permutation group of $Q$ elements weighted by Weingarten functions. The kets $\kket{g_{1,2}}$ are defined in Eq.~\ref{eq:g}. }
\end{figure}

We turn now to the averaged monitored measurements. For a single qubit the average is
given by 
\begin{align}
\mathcal{M} & =\sum_{m}\left(M_{m}\otimes M_{m}^{*}\right)^{\otimes Q} =\left(1-p\right)^{Q}+p^{Q}\sum_{i}\kket{i,\ldots,i;i,\ldots,i}\bbra{i,\ldots,i;i,\ldots,i}.
\end{align}
Finally, for the un-monitored measurements on a single qubit we find 
\begin{align}
\mathcal{K} & =\left(\sum_{\mu}K_{\mu}\otimes K_{\mu}^{*}\right)^{\otimes Q}=\left[\left(1-q\right)+q\sum_{i}\kket{i;i}\bbra{i;i}\right]^{\otimes Q}=\left[\sum_{ij}\left[\left(1-q\right)+q\delta_{i,j}\right]\kket{i;j}\bbra{i;j}\right]^{\otimes Q} \nonumber \\
 & =\sum_{\left[i\right]\left[j\right]}\prod_{k=1}^{Q}\left[\left(1-q\right)+q\delta_{i_{k},j_{k}}\right]\kket{i_{1},\ldots,i_{Q};j_{1},\ldots,j_{Q}}\bbra{i_{1},\ldots,i_{Q};j_{1},\ldots,j_{Q}}
 \label{eq:Kdef}
\end{align}

\subsection{Matrix Elements and Symmetries}

The matrix element between two $g$ states is given by 
\begin{align}
\bbrakket g{g'} & =\sum_{\left[i\right]}\bbrakket{i_{g\left(1\right)},\ldots,i_{g\left(Q\right)}}{i_{g'\left(1\right)},\ldots,i_{g'\left(Q\right)}}\\
 & =\sum_{\left[i\right]}\bbrakket{i_{g'^{-1}g\left(1\right)},\ldots,i_{g'^{-1}g\left(Q\right)}}{i_{1},\ldots,i_{Q}}=d^{\abs{gg'^{-1}}}
 \label{eq:gg'}
\end{align}
and 
\begin{align}
\bbrakket g{i,\ldots,i;i,\ldots,i} & =1.
\end{align}

One can further show that
\begin{align}
\bbrakket{h_{L}gh_{R}^{-1}}{h_{L}g'h_{R}^{-1}} & =\bbrakket g{g'}
\end{align}
so there is a double (left and right) permutation invariance for this matrix element that is of great importance for this discussion. 

It is useful to define operators $\mathcal{P}$ that implements the permutations $h_L$ and $h_R$ and another operator $\mathcal{S}$ that swaps bras and kets (or left and right indices in the vectorized notation)
\begin{align}
\mathcal{P}_{h_{L}h_{R}} & =\sum_{\left[i\right]\left[j\right]}\kket{i_{h_{R}^{-1}\left(1\right)},\ldots,i_{h_{R}^{-1}\left(Q\right)};j_{h_{L}^{-1}\left(1\right)},\ldots,j_{h_{L}^{-1}\left(Q\right)}}\bbra{i_{1},\ldots,i_{Q};j_{1},\ldots,j_{Q}}\\
\mathcal{S} & =\sum_{\left[i\right]\left[j\right]}\kket{j_{1},\ldots,j_{Q};i_{1},\ldots,i_{Q}}\bbra{i_{1},\ldots,i_{Q};j_{1},\ldots,j_{Q}}
\end{align}
One may check that $\mathcal{P}$  acting on state $\kket g$ leads to
\begin{align}
\mathcal{P}_{h_{L}h_{R}}\kket g & =\sum_{\left[i\right]\left[i'\right]}\kket{i_{h_{R}^{-1}\left(1\right)},\ldots,i_{h_{R}^{-1}\left(Q\right)};i'_{h_{L}^{-1}\left(1\right)},\ldots,i'_{h_{L}^{-1}\left(Q\right)}}\bbrakket{i_{1},\ldots,i_{Q}}{i'_{g\left(1\right)},\ldots,i'_{g\left(Q\right)}} \nonumber \\
 & =\sum_{\left[i'\right]}\kket{i'_{gh_{R}^{-1}\left(1\right)},\ldots,i'_{gh_{R}^{-1}\left(Q\right)};i'_{h_{L}^{-1}\left(1\right)},\ldots,i'_{h_{L}^{-1}\left(Q\right)}} \nonumber \\
 & =\sum_{\left[i\right]}\kket{i_{h_{L}gh_{R}^{-1}\left(1\right)},\ldots,i_{h_{L}gh_{R}^{-1}\left(Q\right)};i{}_{1},\ldots,i{}_{1}}=\kket{h_{L}gh_{R}^{-1}},
\end{align}
where in the last equality we perform the change of variables $i'_{h_{L}^{-1}\left(k\right)}=i_{k},i'_{k}=i_{h_{L}\left(k\right)}$.

Similarly, one may look at the action of $\mathcal{S}$ 
\begin{align}
\mathcal{S}\kket g & =\sum_{\left[i\right]}\kket{i_{1},\ldots,i_{Q},i_{g\left(1\right)},\ldots,i_{g\left(Q\right)}}
 =\sum_{\left[i\right]}\kket{i_{g^{-1}\left(1\right)},\ldots,i_{g^{-1}\left(Q\right)},i_{1},\ldots,i_{Q}}=\kket{g^{-1}}.
\end{align}

The matrix element of the projective measurement operator is
\begin{align}
\bbra g\mathcal{M}\kket{g'} & =\left(1-p\right)^{Q}+p^{Q}d.
\end{align}
This operator is also left-/right- permutation symmetric and also symmetric under the swap $\mathcal{S}$
\begin{align}
\mathcal{P}_{h_{L}h_{R}}^{-1}\mathcal{M}\mathcal{P}_{h_{L}h_{R}} & =\mathcal{M},\\
\mathcal{S}^{-1}\mathcal{M}\mathcal{S} & =\mathcal{M}.
\end{align}
Taken together with the average over unitaries it follows that the model in the permutation variables has global symmetries $S_Q \times S_Q \rtimes \mathbb{Z}_2$.

However, when we consider the un-monitored measurements we find
\begin{align}
\mathcal{P}_{h_{L}h_{R}}^{-1}\mathcal{K}\mathcal{P}_{h_{L}h_{R}} & \neq\mathcal{K}\text{ in general for }\ h_{L}\neq h_{R},\\
\mathcal{P}_{hh}^{-1}\mathcal{K}\mathcal{P}_{hh} & =\mathcal{K},\\
\mathcal{S}^{-1}\mathcal{K}\mathcal{S} & =\mathcal{K}
\end{align}
so there is only one surviving permutation invariance $-$ the one where left- and right- are locked to one another. 

\subsection{The Un-Monitored Measurements}

It remains to evaluate the matrix elements for the combined monitored and un-monitored measurements. 

The action of $\mathcal{M}$ on a permutation degree of freedom is $g'$ is
\begin{align}
\mathcal{M}\kket{g'} & =\left[\left(1-p\right)^{Q}\kket{g'}+p^{Q}\sum_{i}\kket{i,\ldots,i;i,\ldots,i}\right].
\end{align}
We note that
\begin{align}
\mathcal{K}\kket{i,\ldots,i;i,\ldots,i} & =\kket{i,\ldots,i;i,\ldots,i}.
\end{align}

Now acting with $\mathcal{K}$ leads to
\begin{align}
\bbra g\mathcal{K}\mathcal{M}\kket{g'} & =\bbra g\mathcal{K}\left[\left(1-p\right)^{Q}\kket{g'}+p^{Q}\sum_{i}\kket{i,\ldots,i;i,\ldots,i}\right] \nonumber \\
 & =\left[\left(1-p\right)^{Q}\bbra g\mathcal{K}\kket{g'}+p^{Q}\sum_{i}\bbra g\mathcal{K}\kket{i,\ldots,i;i,\ldots,i}\right] \nonumber \\
 & =\left(1-p\right)^{Q}\bbra g\mathcal{K}\kket{g'}+p^{Q}d.
 \label{eq:gKMg}
\end{align}

Our goal now is to compute the matrix elements $\bbra g\mathcal{K}\kket{g'}$ appearing in \eqref{eq:gKMg}. The operator $\mathcal{K}$ is given by \eqref{eq:Kdef}.
\begin{align}
\bbra g\mathcal{K}\kket{g'}= & \bbra g\left(\sum_{\left[i\right]\left[j\right]}\prod_{k=1}^{Q}\left[\left(1-q\right)+q\delta_{i_{k},j_{k}}\right]\kket{i_{1}...i_{Q};j_{1}...j_{Q}}\bbra{i_{1}...i_{Q};j_{1}...j_{Q}}\right)\kket{g'}\\
= & \sum_{\left[i\right]\left[j\right]}\prod_{k=1}^{Q}\left[\left(1-q\right)+q\delta_{i_{k},j_{k}}\right]\bbrakket{j_{g\left(1\right)},...,j_{g\left(Q\right)}}{i_{1}...i_{Q}}\bbrakket{i_{1}...i_{Q}}{j_{g'\left(1\right)},...,j_{g'\left(Q\right)}}\\
    = & \left((1-q)^{Q}\sum_{\left[i\right]\left[j\right]}+\sum_{l=1}^{Q}q^{l}(1-q)^{Q-l}\sum_{\{r_{1},\ldots,r_{l}\}\in B_{l,Q}}\sum_{\left[i\right]\left[j\right]}\delta_{i_{r_{1}},j_{r_{1}}}\ldots\delta_{i_{r_{l}},j_{r_{l}}}\right) \nonumber \\
    &\bbrakket{j_{g\left(1\right)},...,j_{g\left(Q\right)}}{i_{1}...i_{Q}}\bbrakket{i_{1}...i_{Q}}{j_{g'\left(1\right)},...,j_{g'\left(Q\right)}}
\end{align}
where $B_{l,Q}$ is the set of subsets of $\{1,2,\ldots,Q-1,Q\}$ with
$l$ elements.

The first term corresponds to performing no un-monitored measurements.
As seen in \eqref{eq:gg'}, this yields 
\begin{align}
    &\sum_{\left[i\right]\left[j\right]}\bbrakket{j_{g\left(1\right)},...,j_{g\left(Q\right)}}{i_{1}...i_{Q}}\bbrakket{i_{1}...i_{Q}}{j_{g'\left(1\right)},...,j_{g'\left(Q\right)}}  \nonumber\\
    &=\sum_{\left[j\right]}\bbrakket{j_{g\left(1\right)},...,j_{g\left(Q\right)}}{j_{g'\left(1\right)},...,j_{g'\left(Q\right)}}=\bbrakket g{g'}=d^{\abs{gg'^{-1}}}.
\end{align}

The second term corresponds to measuring $l=1,\ldots,Q$ out of the
$Q$ replicas. For a fixed $l$, we sum over the $\binom{Q}{l}$ ways
of choosing which $l$ copies to measure. The coefficient of the $q^{l}(1-q)^{Q-l}$
term is given by
\begin{align}
\sum_{\left[i\right]\left[j\right]}\sum_{\{r_{1},\ldots,r_{l}\}\in B_{l,Q}}\delta_{i_{r_{1}},j_{r_{1}}}\ldots\delta_{i_{r_{l}},j_{r_{l}}}\bbrakket{j_{g\left(1\right)},...,j_{g\left(Q\right)}}{i_{1}...i_{Q}}\bbrakket{i_{1}...i_{Q}}{j_{g'\left(1\right)},...,j_{g'\left(Q\right)}}=d^{|gg'^{-1}h_{r_{1}}^{t_{1}}\ldots h_{r_{l}}^{t_{l}}|}
\end{align}
where $B_{l,Q}$ is the set of subsets of ${1,2,\ldots,Q}$ with $l$ elements, 
\begin{align}
    h_{r}=(g^{-1}(r)\ r)
\end{align}
is a two-cycle that permutes replica index $r$ and $g^{-1}(r)$ and
%\begin{align}
%    t_{k}=gg'^{-1}(e-h_{k})\frac{1+\sign(gg'^{-1}(e-h_{k}))}{2}
%\end{align} 
\begin{align}
    t_{k}=\frac{1+\sign(\vert gg'^{-1}\vert-\vert gg'^{-1}h_{r_k}\vert ) }{2}
\end{align} 
is either 0 or 1. If in the permutation
$gg'^{-1}$, the elements $g^{-1}(r)$ and $r$ belong to different
cycles, then $gg'^{-1}h_{r_{k}}^{t_{k}}$ joins the cycles to which
$g^{-1}(r)$ and $r$ belong and $\abs{gg'^{-1}}-|gg'^{-1}h_{r_{k}}|=1$.
If the elements $g^{-1}(r)$ and $r$ already belong to the same cycle,
then $gg'^{-1}h_{r_{k}}$ separates it such that $g^{-1}(k)$
and $k$ belong to different cycles and $\abs{gg'^{-1}}-|gg'^{-1}h_{r_{k}}|=-1$. In this case, $t_k=0$ and  $gg'^{-1}h_{r_{k}}^{t_k}=gg'^{-1}$. 
Thus, expression $|gg'^{-1}h_{r_{k}}^{t_{k}}|$ selects the permutation
$gg'^{-1}$or $gg'^{-1}h_{r_{k}}$ with fewer cycles, i.e.
\begin{align}
|gg'^{-1}h_{r_{k}}^{t_{k}}|=\begin{cases}
\abs{gg'^{-1}} & {\rm if}\ \abs{gg'^{-1}}=|gg'^{-1}h_{r_{k}}|-1\\
|gg'^{-1}h_{r_{k}}| & {\rm if}\ \abs{gg'^{-1}}=|gg'^{-1}h_{r_{k}}|+1
\end{cases},
\end{align}
where $\abs{gg'^{-1}}=|gg'^{-1}h_{r_{k}}|\pm1$ always. We consider
the null power of a cycle to be the identity cycle. 

Finally,
\begin{align}
\bbra g\mathcal{K}\kket{g'}=(1-q)^{Q}d^{\abs{gg'^{-1}}}+\sum_{l=1}^{Q}q^{l}(1-q)^{Q-l}\sum_{\{r_{1},\ldots,r_{l}\}\in B_{l,Q}}d^{|gg'^{-1}h_{r_{1}}^{t_{1}}\ldots h_{r_{l}}^{t_{l}}|}.
\end{align}
Notice that, setting $q=0$, we recover the expression for monitored measurements alone, given by Eq.~13 in \cite{VasseurLudwig_PRB2020_classicalSMmapping}.

\subsection{Bulk and Boundary contributions}

The entanglement entropy is obtained as the analytic continuation of the difference of factors $\mathcal{Z}_A$ and $\mathcal{Z}_\emptyset$ in Eq.~7 in the main text. For a set of sites in region $A$, $\mathcal{Z}_A$ can be interpreted as a classical partition function with boundary conditions imposed on the set of sites with $t=0$, $\partial_0$, and on the set of sites with $t=T$, $\partial_T$. After applying the rules derived in the last sections, the bulk degrees of freedom are found to be distributed on the anisotropic honeycomb lattice depicted in Fig.~\ref{fig:honeycomb}. Vertical bonds (red), $\mathcal{E}_{\rm V}$, correspond to Weingarten weights and the zigzag bonds (blue), $\mathcal{E}_{\rm ZZ}$, to the matrix elements in Eq.~\ref{eq:gKMg}. 

Assuming the starting state is the high-temperature state, i.e. proportional to the identity, the contribution from the $t=0$ boundary is simply given by $\prod_{i \in \partial_{0}}\bbrakket{g_i}{1}/d^Q$. As the protocol assumes a steady-state is reached much before any of the observables are considered, this boundary condition does not influence any of the following results and may be safely ignored. 

On the contrary, the contributions from the $t=T$ boundary are nontrivial and differ for sites belonging to the set $A$ or to its complement, $B$. This boundary contribution, given by Eq.~\ref{eq:Z_A} can be most conveniently obtained by rewriting Eq.~\ref{eq:SgA} as $ \kket{\Sigma_{{A}}^{\left(n,k\right)}} = \otimes_{i\in B} \kket{1} \otimes_{i\in A} \kket{ g(n,k,1)} $, with $\kket{ g(n,k,1)}$ a double-ket state of the from (\ref{eq:g}). Thus, sites in $\bar A$, contribute with $\prod_{i \in \partial_{T} \cap B }\bbrakket{1}{g_i}$.  A depiction, $g(n,k,1)$, acting on  $Q=kn +1 $ replicas of a single site is given in Fig.~ \ref{fig:permutation}. It is composed of $k$ cyclic permutations, $g_\text{cyc}^n$, of $n$ replica indices acting identically on $k$ sets. The last index is left invariant.
Thus, the contribution of sites in $A$ is given by $\prod_{i \in \partial_{T} \cap A  }\bbrakket{g(n,k,1)}{g_i}$.  

The boundary conditions of the classical model are sketched in Fig.~\ref{fig:boundary}.

\begin{figure}[H]
\centering
\includegraphics[width=0.5\columnwidth]{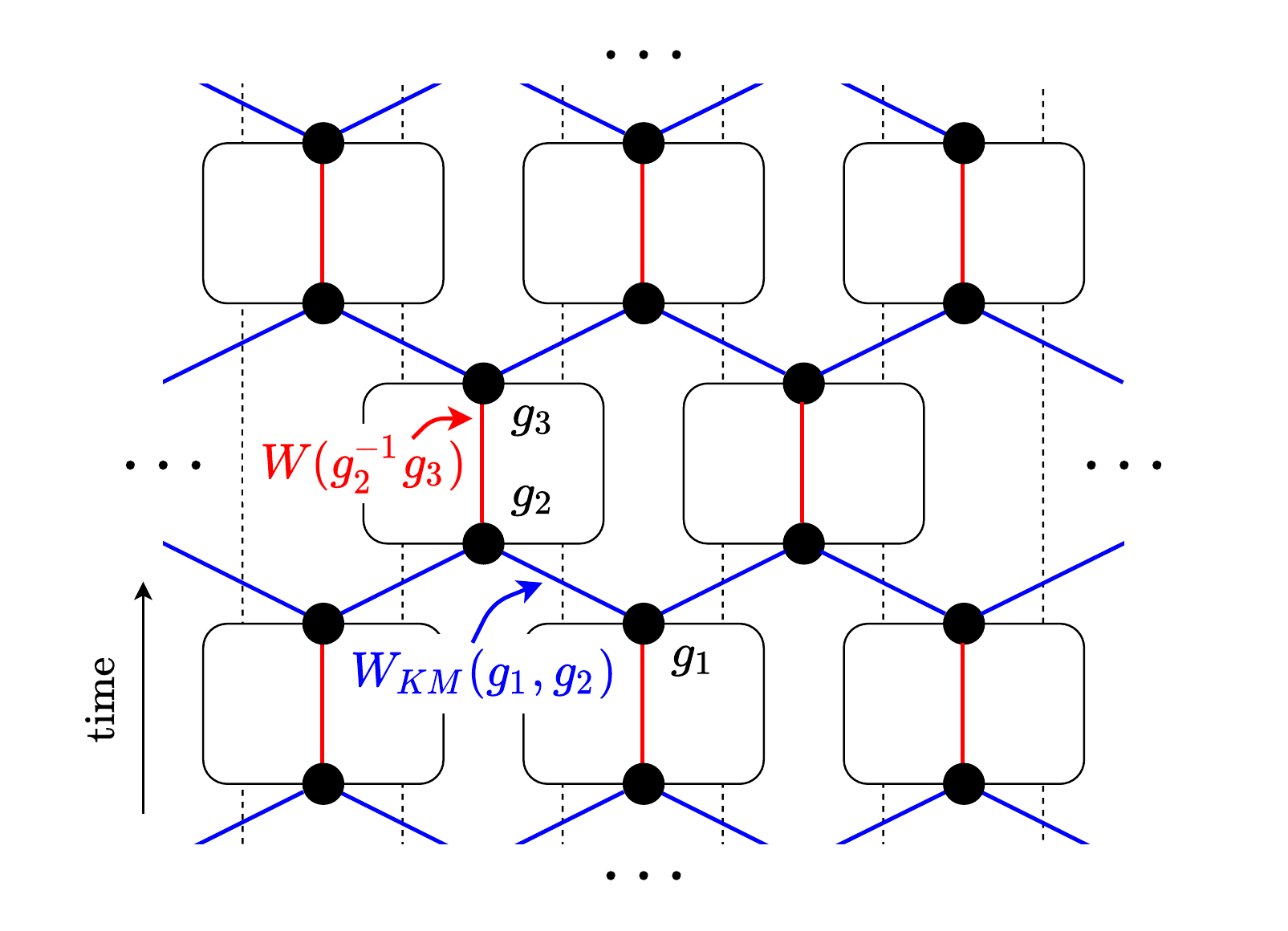}
\caption{\label{fig:honeycomb} Bulk couplings of the effective classical model defined in an anisotropic honeycomb lattice. }
\end{figure}

\begin{figure}[H]
\centering
\includegraphics[width=0.5\columnwidth]{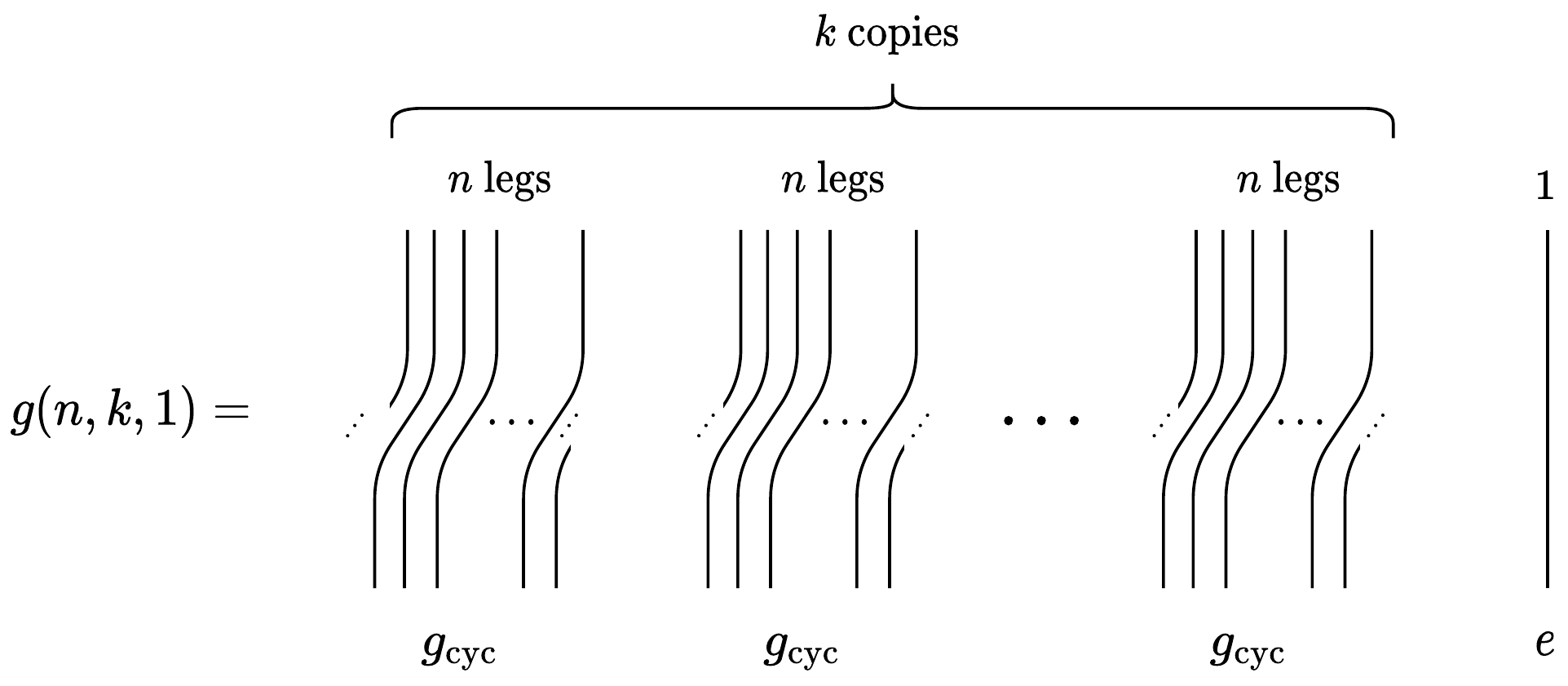}
\caption{\label{fig:permutation} 
$t=T$ boundary permutation, $g(n,k,1)$, acting the $Q=kn +1 $ replica indices for  sites in $A$. $g(n,k,1)$ is composed of $k$ identical cyclic permutations, $g_\text{cyc}^n$, acting on $n$ replica indices. The last index is left invariant. 
}
\end{figure}

\begin{figure}[H]
\centering
\includegraphics[width=0.5\columnwidth]{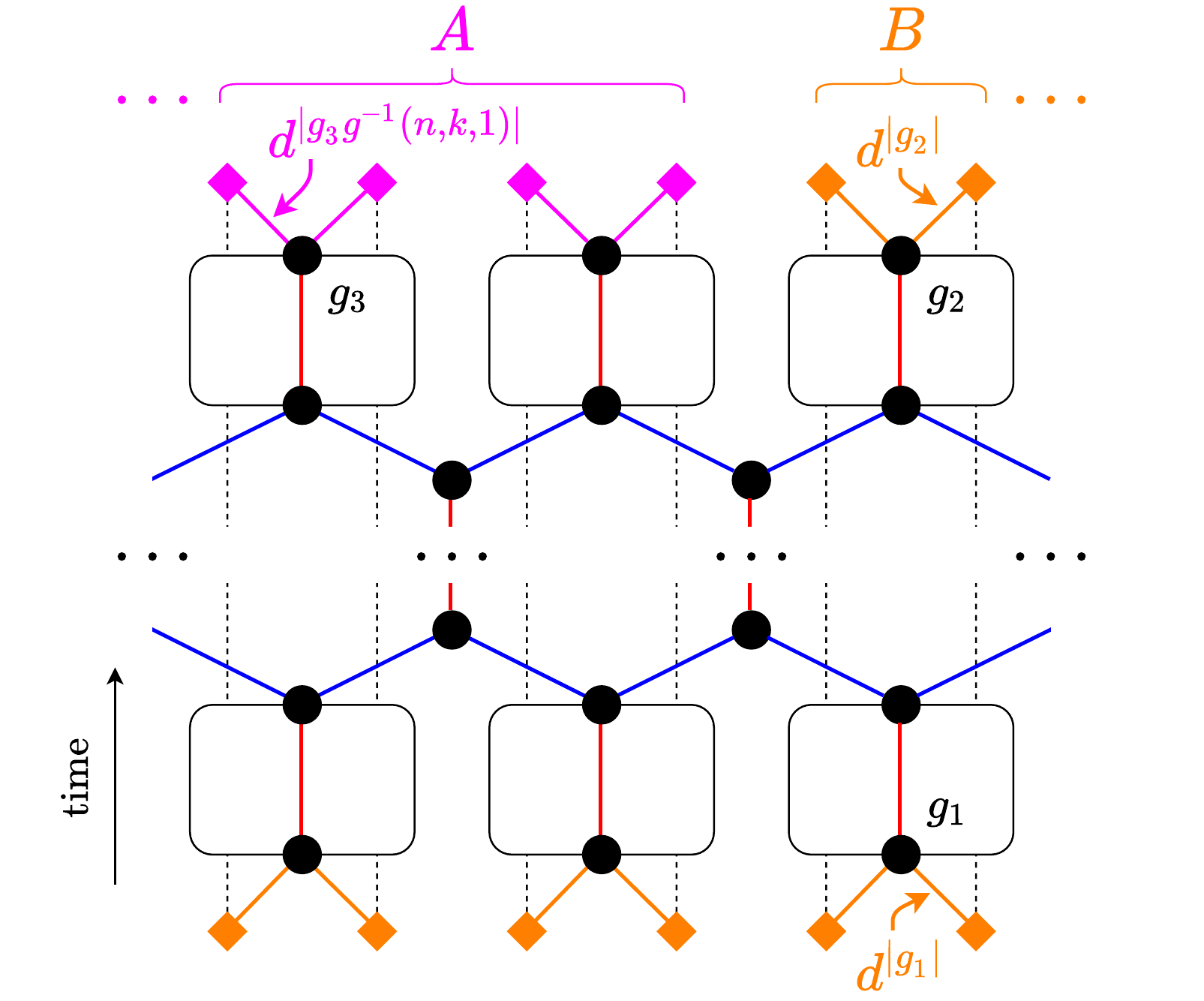}
\caption{\label{fig:boundary} Boundary conditions for the entanglement entropy on subsystem $A$. }
\end{figure}

\subsection{Classical Statistical Model}

So far we have evaluated matrix elements for the measurements on a single qubit and the average over unitaries on a pair of sites. We now use these results to write down the partition function in Eq.~\ref{eq:SgA}: 

\begin{align}
\mathcal{Z}_{A} = \sum_{\left\{g_i \in S_Q\right\}} \left[ \prod_{\langle i,j \rangle \in \mathcal{E}_{\rm V}} W(g_i g_j^{-1}) \times \prod_{\langle i,j \rangle \in \mathcal{E}_{\rm ZZ}} W_{KM}(g_i, g_j) \times \prod_{i\in \partial_0} d^{\abs{g_i}} \times  \prod_{i\in \partial_T \cap B } d^{\abs{g_i}} \times  \prod_{i\in \partial_T \cap A } d^{\abs{g(n,k,1)^{-1} g_i}} \right]
\end{align}
where we reproduce here for convenience the weights computed above
\begin{align}
W_{KM}(g_i, g_j) = \left(1-p\right)^{Q}\bbra{g_i} \mathcal{K}\kket{g_j}+p^{Q}d,
\end{align}
and
\begin{align}
\bbra{g_i} \mathcal{K}\kket{g_j}=(1-q)^{Q}d^{\abs{g_i g_j^{-1}}}+\sum_{l=1}^{Q}q^{l}(1-q)^{Q-l}\sum_{\{r_{1},\ldots,r_{l}\}\in B_{l,Q}}d^{|g_i g_j^{-1}h_{r_{1}}^{t_{1}}\ldots h_{r_{l}}^{t_{l}}|}.
\end{align}
This reduces to the expression in Ref.~\cite{VasseurLudwig_PRB2020_classicalSMmapping} for $q=0$ whereupon $\bbra{g_i} \mathcal{K}\kket{g_j}= d^{\abs{g_i g_j^{-1}}}$. In that case the weights may be interpreted as follows: when $p$ is large the weight is nearly invariant under changes to the states corresponding to the paramagnetic regime and, for small $p$, the system tries to maximize the number of cycles in $g_i g_j^{-1}$ corresponding to an identity permutation where the permutation degrees of freedom are ferromagnetic spontaneously breaking one permutation symmetry.

% As the unitaries and measurements scramble the state all memory of the initial state is expected to be washed out and the partition function is

In the presence of $q\neq 0$, the symmetry that is spontaneously broken for small $p$ is explicitly broken by $q$ analogous to a field applied to a classical ferromagnet. This is consistent with the crossover observed numerically.  

\subsection{Limiting cases}

We consider limiting cases that serve to illustrate some features of the model. We first linearize in $q$:
\begin{align}
\bbra g\mathcal{K}\kket{g'}=(1-q)d^{\abs{gg'^{-1}}} + 
q \sum_{r=1}^Qd^{|gg'^{-1}h_{r}^{t}|} + \mathcal{O}(q^2).
\label{eq:Kggq0}
\end{align}
We then take the $d \rightarrow \infty$ limit. For that, let us take \eqref{eq:Kggq0} up to $\mathcal{O}(d^{Q-2})$.
We have
\begin{align}
d^{\abs{gg'^{-1}}} = \sum_{k=1}^{Q} d^k \delta_{|g g'^{-1}|, k} = d^Q \delta_{g^{-1},g'^{-1}} + d^{Q-1} \delta_{|g g'^{-1}|,Q-1} + \mathcal{O}(d^{Q-2})
\end{align}
and
\begin{align}
\sum_{r=1}^Qd^{|gg'^{-1}h_{r}^{t}|} &= 
|g|_1 \delta_{g^{-1}, g'^{-1}} d^Q +
\left( \max(|g|_1,|g'^{-1}|_1)\,  \delta_{|g g'^{-1}|, Q-1}
+ Q \delta_{|g g'^{-1}|, Q}\, \delta_{|g|_1, 0} 
\right) d^{Q-1}
+ \mathcal{O}(d^{Q-2}) \nonumber \\
&= |g|_1 \delta_{g^{-1}, g'^{-1}} d^Q +
\left( \max(|g|_1,|g'^{-1}|_1)\,  \delta_{|g g'^{-1}|, Q-1}
+ Q \delta_{g^{-1}, g'^{-1}}\, \delta_{|g|_1, 0} 
\right) d^{Q-1}
+ \mathcal{O}(d^{Q-2})
\end{align}
where we used $\delta_{|g g'^{-1}|,Q} = \delta_{g^{-1},g'^{-1}}$ and where $|g|_1$ denotes the number of 1-cycles in the permutation $g$.
Thus, 
\begin{align}
\bbra g\mathcal{K}\kket{g'} &=
\left((1-q) + q |g|_1 \right) \delta_{g^{-1}, g'^{-1}} d^Q \nonumber \\
&+
\left([1 + q \max(|g|_1,|g'^{-1}|_1) ] \delta_{|g g'^{-1}|, Q-1}
+ q Q \delta_{g^{-1}, g'^{-1}}\, \delta_{|g|_1, 0} \right)  d^{Q-1} + \mathcal{O}(d^{Q-2})
\label{eq:Kggq02}
\end{align}

The leading order in $d$ result in \eqref{eq:Kggq02} reveals that the full $S_Q \times S_Q \rtimes \mathbb{Z}_2$ symmetry of the $q=0$ model is broken down to $S_Q \rtimes \mathbb{Z}_2$.

\end{widetext}

\end{document}